\documentclass[aps,pra,twocolumn,longbibliography,superscriptaddress,floatfix,nobalancelastpage,raggedbottom]{revtex4-1}

\usepackage[T1]{fontenc}
\usepackage{graphicx}  
\usepackage{dcolumn}   
\usepackage{bm}        
\usepackage{amssymb}   
\usepackage{mathtools}
\usepackage{amsmath}
\usepackage{amsthm}
\usepackage{ragged2e}
\usepackage{verbatim}
\usepackage[colorlinks=true,linkcolor=blue,citecolor=blue,urlcolor=blue,plainpages=false,pdfpagelabels]{hyperref}
\usepackage{color,xcolor,colortbl}
\usepackage{multirow}
\usepackage{bbm}
\usepackage{lipsum}
\usepackage{array}
\usepackage{etoolbox}

\usepackage{times}
\usepackage{txfonts}
\usepackage[mathcal]{euscript}

\DeclarePairedDelimiter{\ceil}{\lceil}{\rceil}
\DeclarePairedDelimiter{\floor}{\lfloor}{\rfloor}

\newcommand{\e}{\text{e}}

\newtheorem{theorem}{Theorem}

\allowdisplaybreaks

\makeatletter
\g@addto@macro\bfseries{\boldmath}
\makeatother

\begin{document}

\widetext

\title{Practical figures of merit and thresholds for entanglement distribution in quantum networks}

\author{Sumeet Khatri}\email{skhatr5@lsu.edu} \affiliation{Hearne Institute for Theoretical Physics, Department of Physics and Astronomy, Louisiana State University, Baton Rouge, Louisiana, 70803, USA}
\author{Corey T. Matyas}\affiliation{Hearne Institute for Theoretical Physics, Department of Physics and Astronomy, Louisiana State University, Baton Rouge, Louisiana, 70803, USA}
\author{Aliza U. Siddiqui}\affiliation{Hearne Institute for Theoretical Physics, Department of Physics and Astronomy, Louisiana State University, Baton Rouge, Louisiana, 70803, USA}
\author{Jonathan P. Dowling}\affiliation{Hearne Institute for Theoretical Physics, Department of Physics and Astronomy, Louisiana State University, Baton Rouge, Louisiana, 70803, USA}\affiliation{National Institute of Information and Communications Technology, 4-2-1, Nukui-Kitamachi, Koganei, Tokyo 184-8795, Japan}\affiliation{NYU-ECNU Institute of Physics at NYU Shanghai, Shanghai 200062, China}\affiliation{CAS-Alibaba Quantum Computing Laboratory, USTC, Shanghai 201315, China}   

\date{\today}

\begin{abstract}

	Before global-scale quantum networks become operational, it is important to consider how to evaluate their performance so that they can be built to achieve the desired performance. We propose two practical figures of merit for the performance of a quantum network: the average connection time and the average largest entanglement cluster size. These quantities are based on the generation of elementary links in a quantum network, which is a crucial initial requirement that must be met before any long-range entanglement distribution can be achieved and is inherently probabilistic with current implementations. We obtain bounds on these figures of merit for a particular class of quantum repeater protocols consisting of repeat-until-success elementary link generation followed by joining measurements at intermediate nodes that extend the entanglement range. Our results lead to requirements on quantum memory coherence times, requirements on repeater chain lengths in order to surpass the repeaterless rate limit, and requirements on other aspects of quantum network implementations. These requirements are based solely on the inherently probabilistic nature of elementary link generation in quantum networks, and they apply to networks with arbitrary topology.

\end{abstract}

\maketitle

	
\section{Introduction}
	
	Progress is being made on building the quantum internet \cite{Kim08,Sim17,Cast18,WEH18}, with networks consisting of a handful of nodes currently being developed \cite{HKM+18}. The promise of a quantum internet is the ability to perform quantum information processing tasks, such as quantum teleportation \cite{BBC+93,BFK00}, quantum key distribution \cite{BB84,Eke91,GRG+02,SBPC+09}, quantum clock synchronization \cite{JADW00,UD02,ITDB17}, distributed quantum computation \cite{CEHM99}, and distributed quantum metrology and sensing \cite{KKB+14,DRC17,ZZS18,EFM+18,PKD18,XZCZ19}, on a global scale. 
	
	One of the most common methods for creating long-distance entangled links is to transmit photonic qubits through either free space or optical fibers \cite{SSR+11,VanMeter_book}. However, each of these media is lossy, and the probability of successfully transmitting a photon decays exponentially with the distance between the end points \cite{SveltoBook,KJK_book}. Other sources of loss, such as source and detector inefficiencies, as well as read/write inefficiencies of quantum memories, ultimately make the task of establishing links in a quantum network with photonic qubits inherently probabilistic.
	 
	Quantum repeaters \cite{BDC98,DBC99,SSR+11} can be used to increase the success probability, as well as the fidelity, of long-range entanglement in a quantum network. Several schemes for long-range bipartite and multipartite entanglement distribution in quantum repeater networks have been considered \cite{ZDB12,ZBD16,EKB16,WZM+16,MMG18,ZPD+18,PWD18,DKD18,WPZD19,PD19,GI19b,GI19c}. All of these schemes involve first generating elementary bipartite or multipartite entanglement links and then performing measurements to join the elementary links \footnote{These schemes are distinct from ``one way'' schemes, in which the quantum information to be transmitted is sent directly in encoded form; see Refs. \cite{GKL+03,RHG05,FWH+10,MSD+12,MKL+14,NJKL16,MLK+16,MEL17}. We consider here schemes that involve first generating entanglement and then using that entanglement as a resource to send the desired quantum information.}. In general, both the elementary link generation and the joining measurements are probabilistic. How should we evaluate the performance of these entanglement distribution schemes, and what limits are imposed on them by the probabilistic nature of the operations involved?
	
	In this work, we propose two figures of merit that can be used to evaluate the performance of elementary link generation in quantum networks: the \textit{average connection time} and the \textit{average largest entanglement cluster size}. The average connection time is an important quantity because, as we show, it can be used to calculate rate of entanglement distribution in a network as a function of the elementary link generation probability. The average largest entanglement cluster size is important because it gives an indication of the range over which entanglement distribution can be achieved in practice.
	
	Much work has been devoted to quantifying the performance of quantum repeater networks by using as figures of merit fundamental limits on the rate at which either bipartite or multipartite entanglement and/or a secret key can be generated between points in the network \cite{BCHW15,AML16,LP17,AK17,BA17,CM17,RKB+18,BAKE18,Pir19,Pir19b}. In these works, however, perfect quantum repeaters are assumed, and other practical limitations are not explicitly taken into account.
	
	Both of our figures of merit explicitly take into account the probabilistic generation of elementary links as well as the limited coherence time of quantum memories. We show that they can be used to evaluate the performance of the devices used in an actual quantum network implementation and that they can be used to set device requirements for achieving particular values of the quantities. We do this by obtaining bounds on the two figures of merit for a particular class of quantum repeater protocols based on a repeat-until-success strategy executed on a graph-based network topology. These bounds represent limitations on quantum networks based solely on elementary link generation probabilities, and they are independent of any particular physical platform. They can thus serve as a guide for building a real quantum internet.
	
	\begin{figure*}
		\centering
		\includegraphics[scale=0.90]{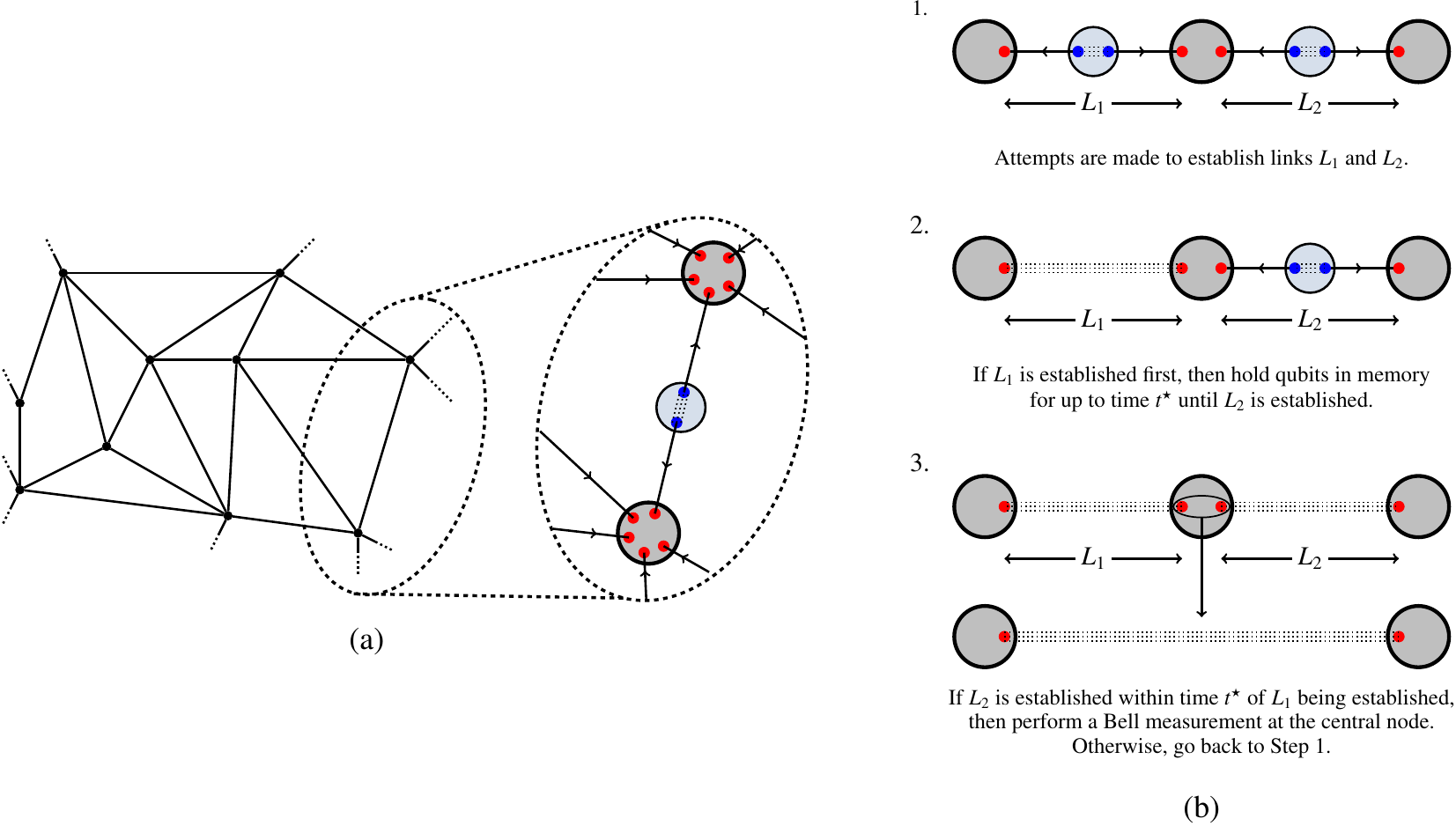}
		\caption{The network architectures that we consider in this work are based on graphs of arbitrary topology. (a) The vertices of the graph correspond to the nodes in the network, and the edges correspond to the elementary links. At the center of each elementary link is a source of entangled photonic qubits (indicated in blue) that fires entangled photons toward the nodes at the ends of the link, where they are held in quantum memories (indicated in red). (b) An example of the general procedure to create bipartite entanglement between two non-adjacent nodes that are connected to a common node.}\label{fig-graph_architecture}
	\end{figure*}
	
	We start in Sec. \ref{sec-architecture_protocol} by outlining the network architecture and quantum repeater protocol that we consider in this work, which generalize the original quantum repeater proposal in Refs. \cite{BDC98,DBC99}. Then, in Sec. \ref{sec-global_dist_time}, we consider the average connection time as a figure of merit and evaluate it for our quantum repeater protocol. We show how the average connection time can be used to compute entanglement distribution rates, and we compare these rates with known repeaterless rate limits. In Sec. \ref{sec-avg_largest_cluster_size}, we consider the average largest entanglement cluster size as a figure of merit for the long-range entanglement distribution capability of a quantum network. We provide concluding remarks in Sec. \ref{sec-summary}.

\section{Network architecture and entanglement distribution protocol}\label{sec-architecture_protocol}

	The network architecture that we consider is illustrated in Fig. \ref{fig-graph_architecture}(a). The network corresponds to an undirected graph $G=(V,E)$, where $V=\{v_i:1\leq i\leq N\}$ is the set of vertices and $E=\{(v_i,v_j):v_i,v_j\in V\}$ is the set of edges. The vertices of the graph correspond to the nodes in the network. The edges of the graph correspond to the elementary links in the network. At the center of each elementary link is a source of bipartite entanglement, which is used to generate bipartite entanglement between the nodes at the ends of the edges. These sources produce entangled photonic qubits in a maximally entangled Bell state. The qubits are encoded into single photons in one of two distinct modes, which are usually horizontal and vertical polarization modes.
	
	The transmission of photons from the source to the neighboring nodes typically occurs through either free space or optical fiber. In each case, loss is the dominant source of noise, which makes the transmission of photons to the nodes probabilistic. In particular, the probability that a photon arrives at a node decreases exponentially with the distance that the photon travels. If we let $\eta_{i,j}$ be the probability that both photons of a pair fired along the edge $(v_i,v_j)$ reach the nodes $v_i$ and $v_j$, then
	\begin{equation}
		\eta_{i,j}=\e^{-\alpha\ell_{i,j}},
	\end{equation}
	where $\ell_{i,j}$ is the length of the edge and $\alpha$ is a value that characterizes the medium. Typically, $\alpha=1/22\text{ km}$ \cite{PRML14}. In the context of dual-rail photonic qubits that we consider here, loss corresponds to an erasure channel between the links (see, e.g., Ref. \cite{BH14,DKD18}), so that with probability $\eta_{i,j}$ both photons arrive at their destination with fidelity unchanged, and with probability $1-\eta_{i,j}$ at least one of the photons is lost, meaning that the state in the corresponding mode is the vacuum state. 
	
	Each node $v_i$ in the network contains $d_i$ quantum memories, where $d_i$ is the degree of the node $v_i$. (The degree of a node is defined to be the number of edges connected to that node.) Several different platforms have been considered for quantum memories in quantum repeater networks, such as trapped ions \cite{SDS09}, Rydberg atoms \cite{ZMHZ10,HHH+10}, atom-cavity systems \cite{LSSH01,RR15}, NV centers in diamond \cite{DMD+13,WCT+14,NTD+16,DHR17,RGR+18,RYG+18}, individual rare-earth ions in crystals \cite{KLW+18}, and superconducting processors \cite{KLS18}. In order to store the arriving photonic qubit state in the quantum memory, each node has locally an optical Bell measurement device. First, a memory-photon entangled state is generated, then a Bell measurement is performed on the photon from the memory-photon pair and the incoming photon from the source. This strategy allows for direct knowledge about the arrival of the photon, which is then communicated to the neighboring node (see, e.g., Ref. \cite{DHR17}). At the same time, conditioned on the success of the Bell measurement, the state of the photonic qubit is transferred to the memory qubit. Linear-optical Bell measurements are limited to a success probability of 50\% \cite{LCS99,CL01,VY99}, although higher success probabilities are in principle possible using nonlinear elements or by increasing the number of photons \cite{KKS01,LSSH01,KKS02,Grice11,WHF+16}.
	
	In addition to loss due to the transmission of photons through free space or optical fibers, there are other sources of loss, such as source inefficiency, detector inefficiency, and quantum memory read/write inefficiency. We can combine all of these loss factors into a single probability $p_{i,j}$ for establishing a link between neighboring nodes $v_i$ and $v_j$. 
	
	To generate elementary links in a network, we use a repeat-until-success strategy in which sources of photonic qubits (blue nodes in Fig. \ref{fig-graph_architecture}) continuously fire entangled states toward repeater stations (gray nodes in Fig. \ref{fig-graph_architecture}) at a rate of $R$ trials per second. Once an elementary link has been established, the corresponding qubits are held in the quantum memories at the repeater nodes for up to time $t^{\star}$ while the neighboring elementary links are being established. After time $t^{\star}$, the effects of decoherence on the stored qubits are considered to be too great and the link is discarded and must be reestablished. The cutoff $t^{\star}$ can take into account not only the coherence times of the quantum memories, but also other more stringent practical requirements. For example, for protocols making use of entanglement purification, the cutoff time should be such that the end-to-end shared entangled states have sufficiently high fidelity in order to perform the desired entanglement purification protocol.
	
	As an example of this repeat-until-success protocol, consider the situation depicted in Fig. \ref{fig-graph_architecture}(b), in which two end nodes are separated via elementary links by a common central node. Generating bipartite entanglement between these end nodes is the most basic element of any long-distance entanglement distribution scheme. Our protocol for establishing entanglement between the end nodes is the following repeat-until-success procedure, which is based on the schemes presented in Refs. \cite{DLCZ01,SRM+07,SSM+07,JKR+16,DHR17,RYG+18}.
	\begin{enumerate}
		\item Elementary link generation attempts are continuously made at a rate of $R$ trials per second. In each trial, a pair of entangled photons is fired from a source station towards the nodes at the ends of the link. Neighboring nodes at the ends of the elementary link communicate classically to confirm the arrival of both photons.
		\item Once an elementary link has been established, the corresponding qubits are stored in quantum memory for up to time $t^{\star}$. If this time is reached and the other elementary link has not been established, then the link must be re-established.
		\item Once both elementary links have been established, an entanglement swapping measurement is performed on the memory qubits in the central node in order to establish entanglement between the end nodes.
	\end{enumerate}

	The protocol described above for generating bipartite entanglement between two nodes separated by one central node generalizes straightforwardly both to bipartite entanglement generation through a longer chain of elementary links and to multipartite entanglement generation over a collection of adjacent elementary links. In these cases, an elementary link must be reestablished after the cutoff time, which means that all of the relevant elementary links must be established before the cutoff of any one of the elementary links is reached. Once all of the relevant elementary links have been established, measurements can be made on the intermediate nodes in order to generate bipartite or multipartite entanglement between the end nodes. Bell measurements are typically used to obtain long-range bipartite entanglement, while multiqubit measurements can be made in order to generate multipartite entanglement; see, e.g., Refs. \cite{ZDB12,ZBD16,EKB16,WZM+16,MMG18,ZPD+18,PWD18,DKD18,WPZD19,PD19,GI19b,GI19c}. In this way, the protocol that we consider is an extension of the original quantum repeater protocol in Refs. \cite{BDC98,DBC99} from a linear chain to an arbitrary topology.
	
\begin{figure}
		\centering
		\includegraphics[width=\columnwidth]{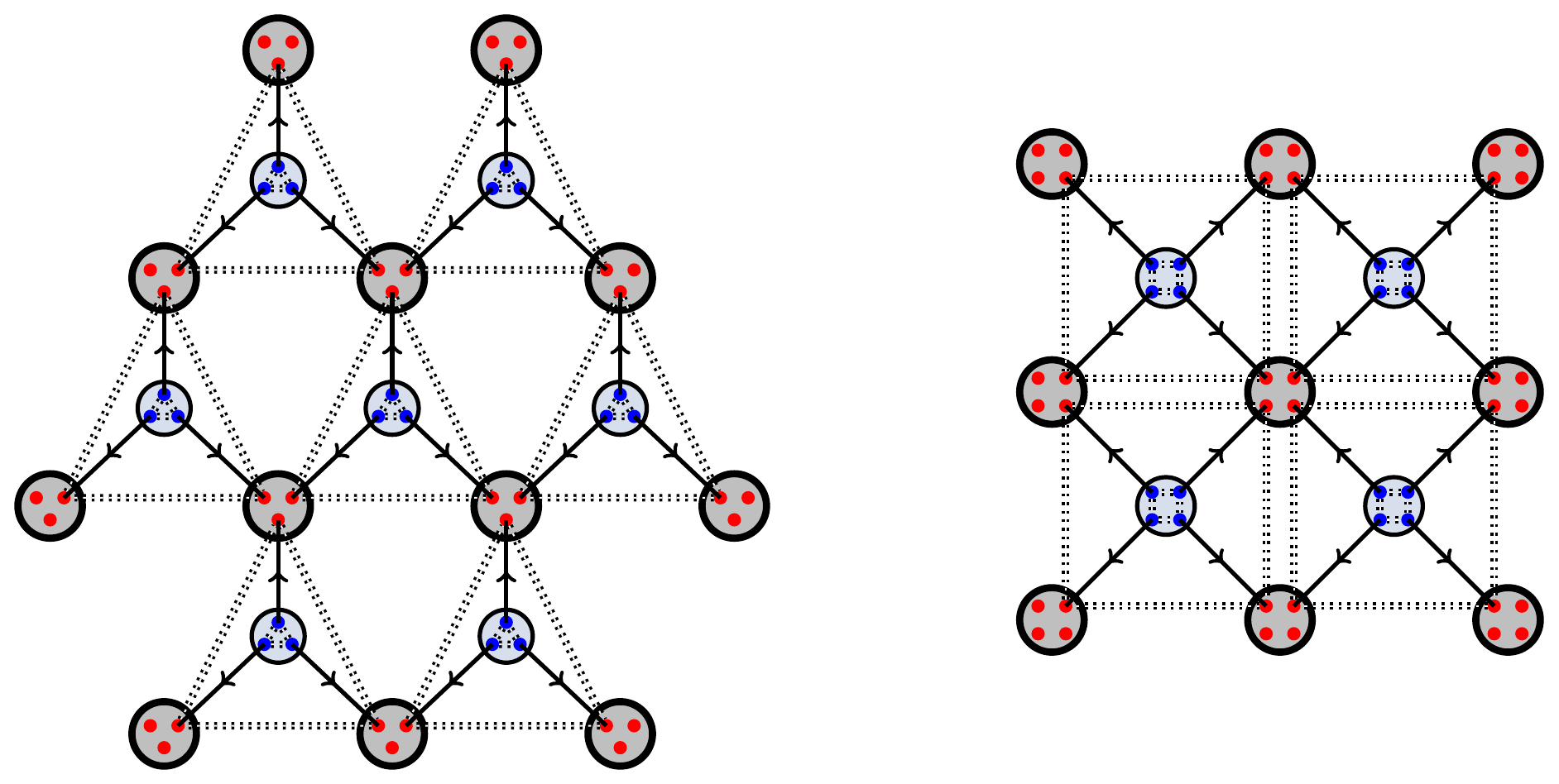}
		\caption{Instead of bipartite entanglement, as in Fig. \ref{fig-graph_architecture}(a), the elementary links in a quantum network can consist of multipartite entanglement; for example, we can have elementary links of tripartite (left) or four-partite (right) entanglement.}\label{fig-gen_architecture_multi}
	\end{figure}

	Another way to generalize the original quantum repeater protocol is to generate elementary links of multipartite entanglement instead of bipartite entanglement. For example, in Fig. \ref{fig-gen_architecture_multi}, the elementary links consist of tripartite entanglement \cite{WZM+16,KVS+19} and four-partite entanglement. One can then consider multipartite entanglement swapping (see, e.g., \cite{WZM+16}) to extend the range of multipartite entanglement. 
	
	The bipartite and multipartite generalizations of the original quantum repeater protocol of Refs. \cite{BDC98,DBC99} that we consider here are similar to the bipartite and multipartite quantum repeater protocols in Refs. \cite{AML16,BA17,AK17,RKB+18,BAKE18,Pir19}. For the figures of merit considered in those works, however, no physical limitations are placed on the quantum repeaters, while we consider the practically relevant scenario of probabilistic elementary link generation and quantum repeaters with limited coherence times.
	
	In our network architecture, we allow for the possibility of having multiple parallel links along the edges connecting two neighboring nodes. This can be achieved using multiple optical fibers between the two nodes, or by employing spectral multiplexing; see, e.g., Refs. \cite{MHS+10,SSM+14,NTD+16}. If $p_{i,j}$ is the probability of establishing a connection along the edge $(v_i,v_j)$ for one of the parallel links, and there are $n>1$ parallel links, then the probability of establishing a connection increases from $p_{i,j}$ to $1-(1-p_{i,j})^n$. In other words, with probability $1-(1-p_{i,j})^n$, at least one of the parallel links successfully connects the two neighboring nodes.
	
	We remark that with nonideal quantum memories, the entanglement distribution protocol described above will in general generate a mixed entangled state between the end nodes with nonunit fidelity to the ideal state. In order to increase the fidelity, one can perform entanglement purification \footnote{See Refs. \cite{BBP96,DAR96,BDD96} for bipartite entanglement purification protocols, and Refs. \cite{DCT99,DC00,DAB03,ADB05,HDB05,MB05} for examples of multipartite entanglement purification protocols. See also Refs. \cite{ZPZ01,YKI01,PSBZ01,DVDV03,BM05,MMO+07,CB08,NFB14,RST+18,KAJ19}.} at the intermediate nodes before performing the measurements that increase the range of entanglement (see, e.g., Refs. \cite{AK17,BA17}). Since entanglement purification protocols are generally probabilistic, the success probability of entanglement purification can be incorporated into the probability $p_{i,j}$ of successfully obtaining an entangled link along the edge $(v_i,v_j)$. Our results thus apply even in the case that entanglement purification between neighboring nodes is incorporated into the entanglement distribution protocol.
	
	A slight modification of the entanglement generation protocol given above is based on the scheme presented in Ref. \cite{BK05}. In this alternate scheme, we place a linear-optical Bell measurement station at the center of each elementary link instead of a source producing photonic-qubit Bell states. An entangled state between a quantum memory and a photon is generated locally at two neighboring nodes. The photon from each node is then transmitted toward the center of the elementary link connecting the two nodes. A Bell measurement is then performed on the two arriving photons. Success of this Bell measurement heralds the generation of entanglement between the two memory qubits at the neighboring nodes. All of the results presented here apply equally to this method. See also \cite{JKR+16,DHR17} for an analysis of this alternative entanglement generation protocol.
	
	To summarize, we consider in this work a quantum repeater protocol in which the elementary link generation is inherently probabilistic. All that matters for our results is the probability $p_{i,j}$ for successfully establishing an elementary link along the edge $(v_i,v_j)$ and the cutoff time $t^{\star}$ of the quantum memories. We do not focus on any particular implementation and the corresponding parameters that may lead to specific values for the probabilities $p_{i,j}$. This allows our results to be completely general and applicable to any practical quantum network implementation.

\section{Average connection time}\label{sec-global_dist_time}

	How long does it take for all of the required elementary links to be established in a quantum network? Given a cutoff time of $t^{\star}$ for the quantum memories, the repetition rate $R$ of the trials in the entanglement distribution protocol, as described in the previous section, leads to a cutoff number of trials $n^{\star}\coloneqq \floor{Rt^{\star}}$, beyond which an elementary link must be reestablished. Then, if there are $M$ elementary links to be established, we let $N(M,n^{\star})$ denote the number of trials needed to establish all $M$ elementary links, so that the required connection time is $T(M,n^{\star})\coloneqq N(M,n^{\star})/R$. Note that $N(M,n^{\star})$ depends only on the number $M$ of elementary links and not on the topology of the network, since all elementary link attempts are independent of each other. We are interested in the \textit{average connection time} $\mathbb{E}\left[T(M,n^{\star})\right]$, which we determine by focusing on the average number $\mathbb{E}\left[N(M,n^{\star})\right]$ of trials.
	
	In Ref. \cite[Eq. (5)]{CJKK07}, it was shown that if all of the elementary links in the network have the same success probability $p$, then
	\begin{equation}
		\mathbb{E}[N(2,n^{\star})]=\frac{3-2p(1-(1-p)^{n^{\star}})-2(1-p)^{n^{\star}}}{2(2-p(1-2(1-p)^{n^{\star}})-2(1-p)^{n^{\star}})}.
	\end{equation}
	For higher values of $M$ in the case $n^{\star}=\infty$, we have that $N(M,\infty)=\max\left\{N_1,N_2,\dotsc,N_M\right\}$, where $N_i$, $1\leq i\leq M$, is a geometric random variable with success probability $p_i$ that indicates the number of trials needed to establish a connection in the $i^{\text{th}}$ elementary link. Indeed, in the case of an infinite cutoff time, once an elementary link is established it is possible to wait as long as required for all of the other links to be established. If, for simplicity, we assume that all of the elementary links in the network have the same success probability $p$, then (see Appendix \ref{app-pf_Nfull_M_nInfty}) $\mathbb{E}\left[N(M,\infty)\right]=\sum_{k=1}^M\binom{M}{k}(-1)^{k+1}(1/(1-(1-p)^k))$.
	
	In the case $n^{\star}<\infty$, the number of trials needed to establish all $M$ elementary links can never be less than the number of trials it takes for any one of the elementary links to be established, meaning that $N(M,n^{\star})\geq N_i$ for all $1\leq i\leq M$. In particular, then, $N(M,n^{\star})\geq\max\left\{N_1,\dotsc,N_M\right\}=N(M,\infty)$, which implies that $\mathbb{E}\left[N(M,n^{\star})\right]\geq \mathbb{E}\left[N(M,\infty)\right]$ for all $n^{\star}\leq\infty$. Furthermore, the most number of trials are required when there are no quantum memories, which is equivalent to setting $n^{\star}=0$. Therefore, $\mathbb{E}\left[N(M,n^{\star})\right]\leq\mathbb{E}\left[N(M,0)\right]$ for all $n^{\star}\geq 0$. In the case $n^{\star}=0$, all of the elementary links have to be established in the same trial, and the probability that this occurs is $p^M$. This means that $\Pr\left[N(M,0)=n\right]=p^M(1-p^M)^{n-1}$. Therefore, $\mathbb{E}\left[N(M,0)\right]=1/p^M$. We thus obtain the following result.
	
	\begin{theorem}\label{thm-dist_time_bounds}
		Consider a quantum network in which the success probability of each of the $M\geq 1$ elementary links is $p$. Then, for all $0\leq n^{\star}\leq \infty$,
		\begin{equation}\label{eq-Nfull_bounds}
			\sum_{k=1}^M\binom{M}{k}\frac{(-1)^{k+1}}{1-(1-p)^k}\leq\mathbb{E}[N(M,n^{\star})]\leq\frac{1}{p^M}.
		\end{equation}
	\end{theorem}
	
	(See Appendix \ref{app-pf_Nfull_M_nInfty} for the proof.) Theorem \ref{thm-dist_time_bounds} gives us a lower bound on the average connection time for \textit{any} network, and it depends only on the number $M$ of elementary links being established in the network, as well as the elementary link success probability $p$.
	
	\begin{figure}
		\centering
		\includegraphics[width=\columnwidth]{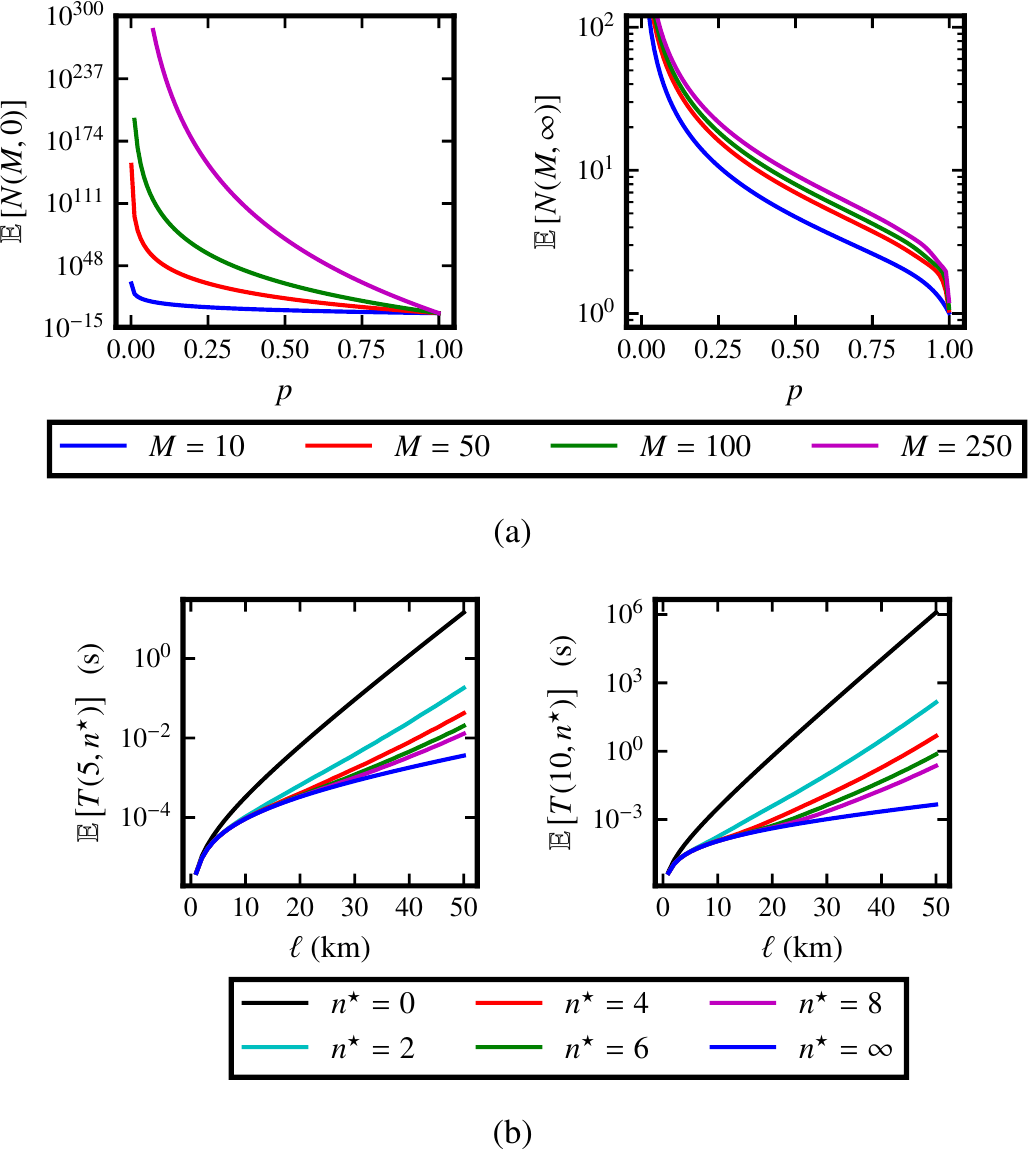}
		\caption{(a) The upper and lower bounds $\mathbb{E}\left[N(M,0)\right]$ and $\mathbb{E}\left[N(M,\infty)\right]$, respectively, from Theorem \ref{thm-dist_time_bounds}, for networks with $M=10,50,100,250$ elementary links. (b) Average connection times $\mathbb{E}\left[T(M,n^{\star})\right]=\mathbb{E}\left[N(M,n^{\star})\right]/R$, assuming $p=\e^{-\alpha\ell}$, $\alpha=\frac{1}{22\text{ km}}$, and $R=\frac{c}{\ell}$, for $M=5,10$ elementary links.}\label{fig-Tfull_n0nInfty}
	\end{figure}
	
	See Fig. \ref{fig-Tfull_n0nInfty}(a) for plots of $\mathbb{E}\left[N(M,0)\right]$ and $\mathbb{E}\left[N(M,\infty)\right]$ for various values of $M$. As an example, suppose that we would like to construct a network with $M=10$ elementary links, and we would like the network to be fully connected within 10 trials on average. Then, in the best-case scenario of $n^{\star}=\infty$, we see from Fig. \ref{fig-Tfull_n0nInfty}(a) that we would require a link success probability $p$ of at least 0.25. Also, if we assume that $p=\e^{-\alpha\ell}$ (with $\alpha=\frac{1}{22\text{ km}}$), so that the photon transmission medium is the only source of loss, if we assume that all elementary links have the same length $\ell$, and we let $R=\frac{c}{\ell}$ be the repetition rate \footnote{Note that the repetition rate $R$ is limited by the latency caused by the need to perform two-way classical communication between neighboring nodes. Other factors contribute to the latency, such as the time needed to write into the quantum memory; see, e.g., Ref. \cite{PRML14,JKR+16}.}, where $c$ is the speed of light, then we find that even for an internodal distance of $\ell=40$ km, the average global connection time for a network with $M=100$ elementary links and no quantum memories is approximately $10^{74}$ seconds, which is longer than the age of the universe. On the other hand, with quantum memories and a cutoff $n^{\star}=\infty$, the average connection time is less than $10^{-2}$ seconds. Note that $10^{-2}$ seconds is the optimal connection time, meaning that any network with $M=100$ elementary links and an internodal distance of $\ell=40$ km making use of the protocol described in Sec. \ref{sec-architecture_protocol} requires at least $10^{-2}$ seconds to become fully connected. We also find that for a network with $M=10$ elementary links and an internodal distance of $\ell=30$ km, it is not possible to obtain a fully connected network in less than $10^{-3}$ seconds. 
	
	In order to obtain tighter estimates for $\mathbb{E}\left[N(M,n^{\star})\right]$ for $0<n^{\star}<\infty$, we resort to estimating $\mathbb{E}\left[N(M,n^{\star})\right]$ via Monte Carlo simulations. Assuming that all elementary links have the same length $\ell$, letting $R=\frac{c}{\ell}$ be the repetition rate as before, and assuming $p=\e^{-\alpha\ell}$, $\alpha=\frac{1}{\text{22 km}}$, we obtain the plots in Fig. \ref{fig-Tfull_n0nInfty}(b) for $\mathbb{E}\left[T(M,n^{\star})\right]=\mathbb{E}\left[N(M,n^{\star})\right]/R$ when $M=5,10$. For example, suppose that we have a network with $M=10$ elementary links and we would like to obtain a fully connected network within one second. Then, with quantum memories such that $n^{\star}=2$, we see from Fig. \ref{fig-Tfull_n0nInfty}(b) that the maximum possible internodal distance is $\ell\approx 37$ km, and this maximum distance corresponds to a cutoff time of $t^{\star}=2\cdot 37\text{ km}/c=246$ $\mu$s. More generally, if we impose a particular cutoff time $t^{\star}$, then the maximum internodal distance is $\ell=ct^{\star}/2$, which corresponds to $n^{\star}=2$, and in this case the average connection time is also maximal. By taking higher values of $n^{\star}$, the average connection time can be decreased at the cost of decreasing the maximum internodal distance.
	
	\begin{table}
		\centering
		\begin{tabular}{|c|c||c|c|c|c|c|c|}
			\hline\multicolumn{2}{|c||}{$p$} & 0.01 & 0.03 & 0.05 & 0.1 & 0.3 & 0.5 \\ \hline\hline
			\multirow{2}{*}{$M$} & 10 & 655 & 210 & 125 & 65 & 18 & 9 \\ \cline{2-8}
			& 20 & 745 & 250 & 150 & 70 & 20 & 10 \\ \hline
		\end{tabular}
		\caption{Minimum cutoffs $n_{\text{min}}^{\star}$ beyond which the average number $\mathbb{E}[N(M,n_{\text{min}}^{\star})]$ of trials is approximately within 1\% of the optimal value, which is $\mathbb{E}\left[N(M,\infty)\right]$.}\label{tab-n_star_crit}
	\end{table}
			
	The lower bound in Theorem \ref{thm-dist_time_bounds} imposes a requirement on the cutoff $n^{\star}$ needed in order to obtain the required elementary links in the fewest possible number of trials on average. In Table \ref{tab-n_star_crit}, we show the minimum cutoff, denoted by $n_{\min}^{\star}$, that is required in order to obtain the required elementary links in a number of trials that is within 1\% of the optimal number $\mathbb{E}\left[N(M,\infty)\right]$ of trials. For values of the link success probability $p$ less than 0.1, we require a cutoff on the order of hundreds of trials. If we assume that $p=\e^{-\alpha\ell}$ for all elementary links, and that all elementary links have the same length $\ell$, then $p=0.01$ implies $\ell\approx 100$ km, so that for $M=10$ elementary links the required cutoff time is approximately $t^{\star}=655\cdot\ell/c=0.2$ seconds.

\subsection{Entanglement distribution rates and overcoming the repeaterless limit}

\begin{figure}
		\centering
		\includegraphics[scale=1]{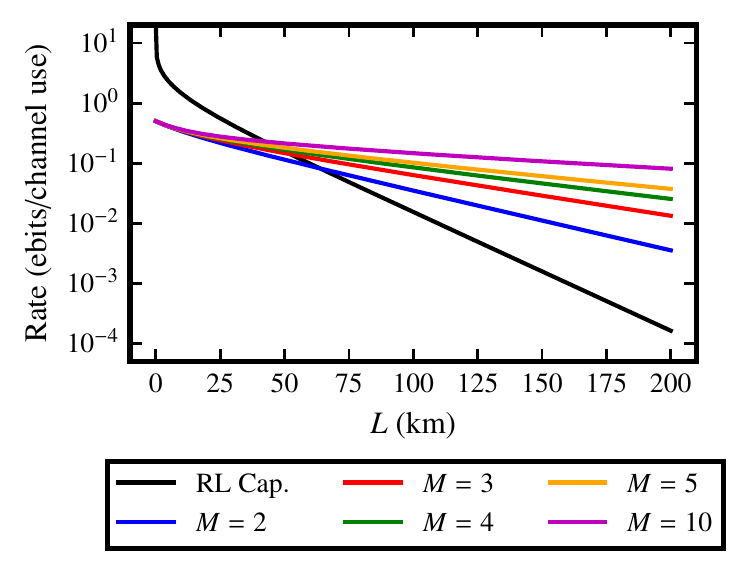}
	 	\caption{Optimal rates $1/(2\mathbb{E}\left[N(M,\infty)\right])$ for the quantum repeater protocols considered in this work, as executed on a linear chain with $M$ elementary links, compared to the repeaterless capacity (RL Cap.) $-\log_2(1-\eta)$ of the bosonic pure-loss channel \cite{PLOB17,WTB17}. The end-to-end distance of the chain is $L$, such that $\eta=\e^{-\alpha L}$, $\alpha=\frac{1}{22\text{ km}}$. To compute the rates $1/(2\mathbb{E}\left[N(M,\infty)\right])$, we set $p=\e^{-\alpha\frac{L}{M}}$.}\label{fig-rates_comp_with_RL}
	 \end{figure}

	A well-established figure of merit for any quantum repeater protocol is the repeaterless (i.e., point-to-point) quantum/secret-key capacity \cite{W18Book} of the quantum communication channel over which the protocol takes place. The quantity $\mathbb{E}\left[N(M,n^{\star})\right]$ can be used to evaluate the quantum repeater protocol that we consider here against this figure of merit. Let us consider a linear repeater chain with a total length $L$ divided into $M$ elementary links, such that there are $M-1$ equally spaced repeater stations between the end points. Photonic qubits are made using the dual-rail encoding. Then, the source stations at the center of each elementary link fire maximally entangled qubit pairs (ebits) towards the repeater stations through a bosonic pure-loss channel \cite{Serafini_book} with transmissivity $\e^{-\alpha\frac{\ell}{2}}$, where $\alpha=\frac{1}{22\text{ km}}$ and $\ell=\frac{L}{M}$ is the length of each elementary link . We assume for simplicity that the entanglement swapping Bell measurements at the repeater stations are perfect and deterministic, and that there are no other imperfections. Note that each trial of the elementary link generation requires two uses of the channel. Furthermore, due to the dual-rail encoding, each trial has success probability $p=\e^{-\alpha\ell}$ \footnote{As a result of the dual-rail encoding, transmission of each mode through the bosonic pure-loss channel corresponds to an erasure channel, in which both photons arrive perfectly with probability $p$, and with probability $1-p$ at least one of the photons is lost, so that the corresponding mode is in the vacuum state; see, e.g., \cite{BH14,DKD18}.}. Therefore, because $\mathbb{E}\left[N(M,n^{\star})\right]$ represents the number of trials needed to obtain one end-to-end ebit, the quantity $1/(2\mathbb{E}\left[N(M,n^{\star})\right])$ is the average number of end-to-end ebits per channel use, i.e., the rate. By Theorem \ref{thm-dist_time_bounds}, $1/(2\mathbb{E}\left[N(M,\infty)\right])$ is the highest possible rate for the protocol that we consider. We compare this rate to the repeaterless capacity of the bosonic pure-loss channel over the entire length of the chain, which is $-\log_2(1-\e^{-\alpha L})$ \cite{PLOB17,WTB17}. The results are shown in Fig. \ref{fig-rates_comp_with_RL}, using which we obtain repeater chain lengths $L_{\min}$ beyond which our repeater protocol can overcome the repeaterless capacity; see Table \ref{tab-threshold_length}. 

	 \begin{table}
	 	\centering
	 	\begin{tabular}{|c||c|c|c|c|c|}
	 		\hline $M$ & 2 & 3 & 4 & 5 & 10 \\ \hline\hline
	 		$L_{\min}$ (km) & 63 & 52 & 47 & 45 & 42 \\ \hline
	 	\end{tabular}
	 	\caption{Minimum total repeater chain lengths $L_{\min}$, as obtained from the results in Fig. \ref{fig-rates_comp_with_RL}, beyond which the quantum repeater protocol considered in this work can overcome the repeaterless capacity of the bosonic pure-loss channel. $M$ is the number of elementary links in the chain.}\label{tab-threshold_length}
	 \end{table}

\subsection{Parallel elementary link generation}

	We can extend Theorem \ref{thm-dist_time_bounds} to the case that $n_P$ parallel paths exist in the network for the required elementary links. The parallel paths can arise either due to multiple parallel transmission channels or through edge-disjoint paths when considering the connection of two distinct points in the network. We let $N(M,n^{\star};n_P)$ denote the number of trials needed to obtain all $M$ elementary links in a network with memory cutoff $n^{\star}$ and $n_P$ parallel paths for the elementary links.
	
	Without quantum memories, i.e., in the case $n^{\star}=0$, $N(M,0;n_P)$ is simply a geometric random variable in which the corresponding success probability $p_{\text{succ}}$ is given by the probability that at least one of the $n_P$ paths has all of its elementary links established. For simplicity, as before, we suppose that each elementary link has a success probability of $p$. Then, $p_{\text{succ}}=1-\left(1-p^{M}\right)^{n_P}$. This holds due to the fact that the $i^{\text{th}}$ path is connected with probability $p^{M}$. Then, with probability $1-p^{M}$, at least one of the elementary links in the $i^{\text{th}}$ path fails. Then, since all paths are independent of each other, the probability that they all fail is $\prod_{i=1}^{n_P}\left(1-p^{M}\right)=\left(1-p^M\right)^{n_P}$. We thus have that
	\begin{equation}\label{eq-N_AB_n0}
		\mathbb{E}\left[N(M,0;n_P)\right]=\frac{1}{\displaystyle 1-\left(1-p^{M}\right)^{n_P}}.
	\end{equation}
	
	Let us now determine $N(M,n^{\star};n_P)$ in the case in which $n^{\star}=\infty$. To start, let $N_j^i$ be the number of trials needed to establish the $j^{\text{th}}$ elementary link in the $i^{\text{th}}$ path, where $1\leq j\leq M$. Furthermore, let $N^i(M,\infty)$ be the number of trials needed to establish the $i^{\text{th}}$ path between $A$ and $B$. Then, the number of trials needed to establish one of the $n_P$ paths between $A$ and $B$ depends on which of the paths gets established first. We thus obtain
	\begin{multline}
		N(M,\infty;n_P)\\=\min\left\{N^1(M,\infty),N^2(M,\infty),\dotsc,N^{n_P}(M,\infty)\right\}.
	\end{multline}
	In Appendix \ref{app-parallel_link}, we show that
	\begin{multline}\label{eq-Pr_NAB_infty}
		\Pr\left[N(M,\infty;n_P)=n\right]\\=\left(1-\left(1-(1-p)^{n-1}\right)^M\right)^{n_P}-\left(1-\left(1-(1-p)^n\right)^M\right)^{n_P},
	\end{multline}
	and that the average number of trials needed to connect the $M$ elementary links is
	\begin{equation}\label{eq-N_AB_nInfty}
		\mathbb{E}\left[N(M,\infty;n_P)\right]=\sum_{n=1}^{\infty}\left(1-\left(1-(1-p)^{n-1}\right)^M\right)^{n_P}.
	\end{equation}
	
	\begin{figure}
		\centering
		\includegraphics[width=\columnwidth]{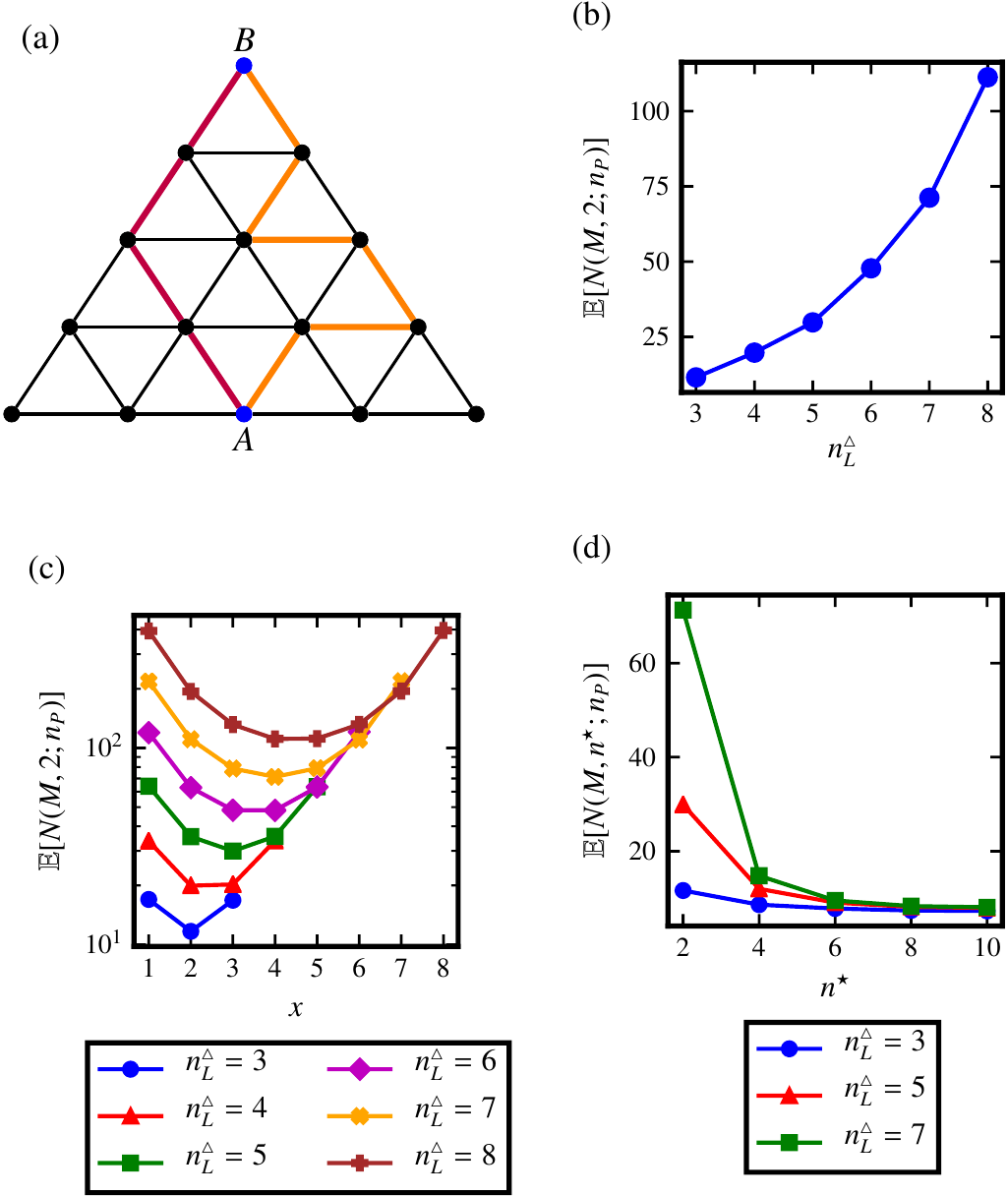}
		\caption{(a) A two-dimensional pyramid network with $n_L^{\bigtriangleup}=5$ layers. Shown are two paths from the node $A$ at the center of the bottom layer to $B$ at the very top of the pyramid. (b) The average number of trials when $A$ is at the center of the bottom layer of the pyramid and $B$ is at the top of the pyramid, as a function of the number $n_L^{\bigtriangleup}$ of layers in the pyramid. We set $n^{\star}=2$ and $p=0.1$. (c) The average number of trials as a function of the position $x$ of $A$ on the bottom layer of the pyramid, with $x=1$ being the left-most corner. The node $B$ is again at the top of the pyramid, and we again set $n^{\star}=2$ and $p=0.1$. (d) The average number of trials as a function of $n^{\star}$ when $A$ is at the center of the bottom layer and $B$ is at the top, with $p=0.1$.}\label{fig-pyramid_AB}
	\end{figure}
	
	Let us now consider the example of a network with the topology of a two-dimensional pyramid, as shown in Fig. \ref{fig-pyramid_AB}(a). We assume that all of the elementary links have the same success probability $p$. We let $n_L^{\bigtriangleup}$ denote the number of layers in the network. 
	
	How many trials does it take, on average, to obtain a connected path from the node at the top of the network to one of the nodes at the bottom? We let $p=0.1$, and we let $A$ be at the center of the bottom layer of the pyramid and $B$ be at the very top of the pyramid. The results we obtain are in Fig. \ref{fig-pyramid_AB}(b) for $n_L^{\bigtriangleup}=3,4,5,6,7,8$ and $n^{\star}=2$. We see that as the size of the network grows, so too does the required number of trials.
	
	Next, we consider the distribution of trials for the nodes on the bottom layer. We again let $p=0.1$ and $n^{\star}=2$. The results we obtain are shown in Fig. \ref{fig-pyramid_AB}(c). The number of trials is symmetric around on the center of the bottom layer for all values of $n_L^{\bigtriangleup}$. In particular, placing $A$ at the center of the bottom layer results in the fewest number of trials, while placing $A$ at either one of the two edges of the bottom layer results in the highest number of trials.
	
	\begin{figure*}
		\centering
		\includegraphics[scale=0.88]{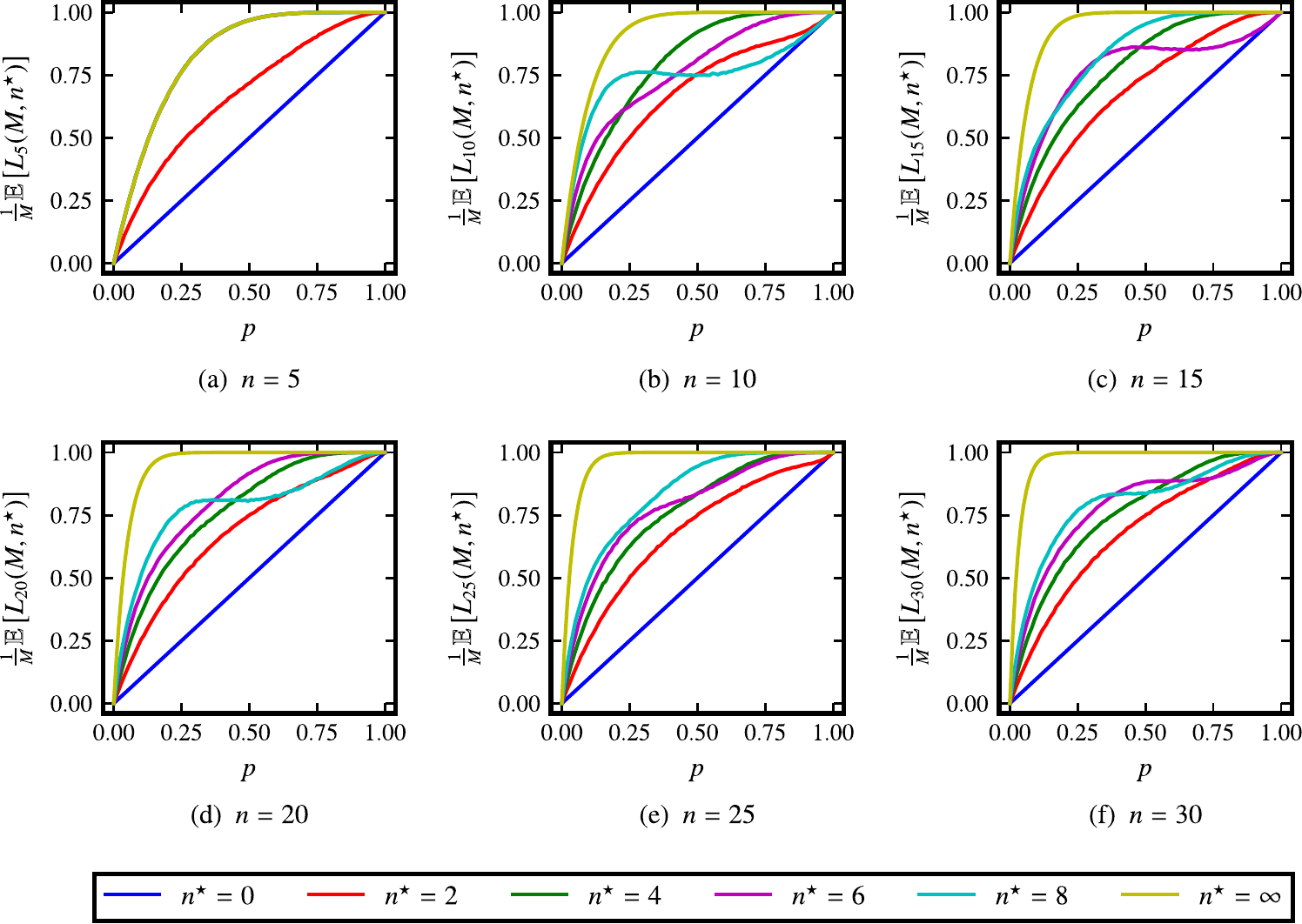}
		\caption{The average fraction $\frac{1}{M}\mathbb{E}\left[L_n(M,n^{\star})\right]$ of established elementary links in a network with $M=40$ elementary links for various values of the cutoff $n^{\star}$. Analytic expressions for $\frac{1}{M}\mathbb{E}\left[L_n(M,0)\right]$ and $\frac{1}{M}\mathbb{E}\left[L_n(M,\infty)\right]$ are given by the left- and right-hand sides, respectively, of Eq. \eqref{eq-L_n_bounds}.}\label{fig-avg_num_links}
	\end{figure*}
	
	Finally, we consider the effect of having longer cutoffs $n^{\star}$. In Fig. \ref{fig-pyramid_AB}(d), we plot the average number of trials needed when $A$ is at the center of the bottom layer and $B$ is at the top, for $n_L^{\bigtriangleup}=3,5,7$ and $p=0.1$. As expected, as $n^{\star}$ increases, the number of trials decreases. Interestingly, for all three network sizes corresponding to $n_L^{\bigtriangleup}=3,5,7$, the average number of trials appears to approach a value close to eight, suggesting that eight is the fewest number of trials in which a connection between $A$ and $B$ can be established, at least for pyramid networks with an odd number of layers.

\section{Average largest entanglement cluster size}\label{sec-avg_largest_cluster_size}

	In order to understand the long-range connectivity in a network, it is important to consider the size of the largest cluster of established elementary links that can be achieved in the network within a certain period of time. By a cluster, we mean a collection of nodes in the network, all of which are connected to each other via established elementary links. We define the size of a cluster by the number of established elementary links contained in it. Since every established elementary link corresponds to a shared entangled state between neighboring nodes, we refer to a cluster as an entanglement cluster.
	
	Let $S_{n}^{\max}(G,n^{\star})$ denote the size of the largest entanglement cluster after $n\geq 1$ trials in a network described by the graph $G=(V,E)$ with memory cutoff $n^{\star}$. We are interested in the quantity $\mathbb{E}\left[S_n^{\max}(G,n^{\star})\right]$, which is the \textit{average largest entanglement cluster size}. Note that the size of the largest entanglement cluster in a network after a certain number of trials can never exceed the number of established elementary links in the entire network after the same number of trials. If we let $L_n(M,n^{\star})$ denote the number of established elementary links after $n$ trials in the network of $M=|E|$ total elementary links, we thus have the upper bound $S_n^{\max}(G,n^{\star})\leq L_n(M,n^{\star})$.
	
	Now, how many elementary links can be established in the network in a fixed amount of time when we start with a network with all elementary links unestablished? As before, we work more generally with the number of trials instead of with the time, and we assume that all elementary links have the same success probability $p$. We are interested in the quantity $\mathbb{E}[L_n(M,n^{\star})]$. Observe that $L_n(M,n^{\star})$ does not depend explicitly on the graph $G$, similarly to the quantity $N(M,n^{\star})$, since all elementary link attempts are independent of each other. We provide more specific details about the quantity $L_n(M,n^{\star})$ in Appendix \ref{app-L_n}.

	With $n^{\star}=0$, all established elementary links are reset after one trial. Then, since all elementary link attempts are independent of each other, we have that $\Pr[L_n(M,0)=x]=\binom{M}{x}p^x(1-p)^{M-x}$, which means that $\frac{1}{M}\mathbb{E}[L_n(M,0)]=p$.
	
	For $n^{\star}>0$, we prove in Appendix \ref{app-L_n} that
	\begin{equation}
		\frac{1}{M}\mathbb{E}[L_n(M,n^{\star})]=1-(1-p)^n,\quad n\leq n^{\star}+1.
	\end{equation}
	In the case $n>n^{\star}+1$, we estimate $\frac{1}{M}\mathbb{E}\left[L_n(M,n^{\star})\right]$ via Monte Carlo simulations. See Fig. \ref{fig-avg_num_links} for plots of $\frac{1}{M}\mathbb{E}\left[L_n(M,n^{\star})\right]$ with $M=40$ for a variety of finite nonzero cutoffs $n^{\star}$. Although we take $M=40$ elementary links in the network to obtain the plots, after comparing the results with various other values of $M$, we find that the fraction $\frac{1}{M}\mathbb{E}\left[L_n(M,n^{\star})\right]$ does not depend on $M$. Furthermore, we find in some cases that having a higher value of $n^{\star}$ is unhelpful for obtaining a higher fraction of established elementary links for certain values of $p$. For example, in the case of $n=30$ trials, we find that for $p$ roughly between 0.30 and 0.70, the average number of established elementary links with $n^{\star}=6$ is higher than the average number of established elementary links with $n^{\star}=8$. This behavior is due to the fact that with a finite cutoff, there are times at which several established elementary links are simultaneously removed as a consequence of reaching the cutoff number of trials, especially when the elementary link success probability $p$ is high. Interestingly, therefore, unlike the quantity $N(M,n^{\star})$, the quantity $L_n(M,n^{\star})$ is not monotonic in $n^{\star}$ for all values of $n$ and $p$.  
	
	In general, the number of established elementary links with a finite cutoff cannot exceed the number of established elementary links with an infinite cutoff. Therefore, $\mathbb{E}[L_n(M,0)]\leq\mathbb{E}[L_n(M,n^{\star})]\leq\mathbb{E}[L_n(M,\infty)]$ for all $0\leq n^{\star}\leq\infty$. Using this, we obtain the following result, the proof of which can be found in Appendix \ref{app-L_n}.
	
	\begin{theorem}\label{thm-L_n_bounds}
		Consider a network described by a graph $G$ with $M$ edges, such that each elementary link has a success probability $p$. Then,
		\begin{equation}\label{eq-L_n_bounds}
			p\leq\frac{1}{M}\mathbb{E}[L_n(M,n^{\star})]\leq 1-(1-p)^n,
		\end{equation}
		and $\frac{1}{M}\mathbb{E}[S_n^{\max}(G,n^{\star})]\leq 1-(1-p)^n$ for all $0\leq n^{\star}\leq\infty$.
	\end{theorem}
	
	As an immediate application of Theorem \ref{thm-L_n_bounds}, suppose that we would like a fraction $f$ of established elementary links in a network with a given elementary link success probability $p$. Then, Theorem \ref{thm-L_n_bounds} tells us that, no matter what the cutoff $n^{\star}$ is, we require at least $n=\ceil{\log(1-f)/\log(1-p)}$ trials on average in order to achieve the desired fraction $f$ of established elementary links.

\begin{figure}
		\centering
		\includegraphics[width=\columnwidth]{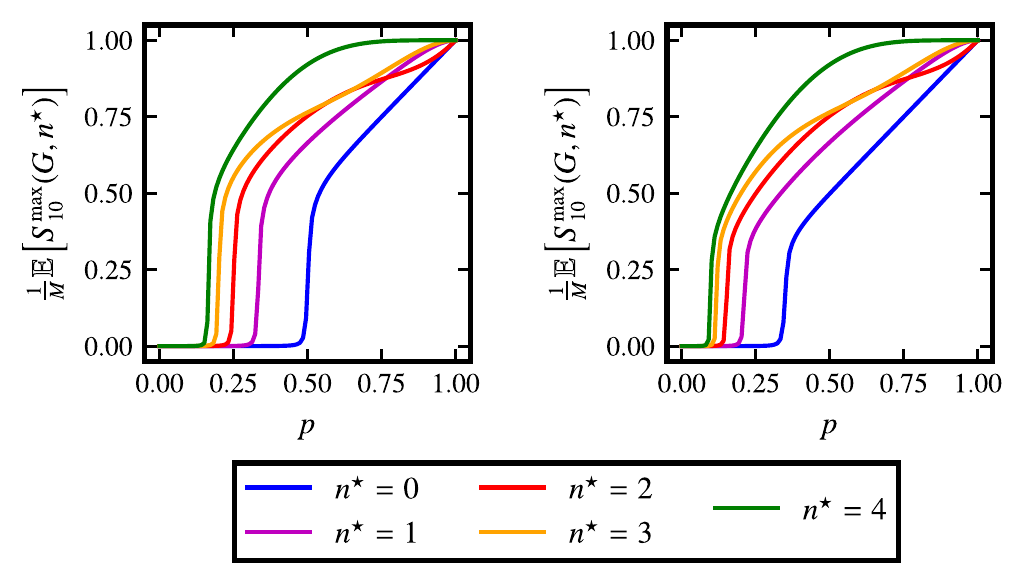}
		\caption{Estimated average largest entanglement cluster sizes with $n=10$ trials and various cutoffs for the $500\times 500$ square (left) and triangular (right) lattices.}\label{fig-avg_cluster_size_all}
	\end{figure}

	Let us now return to the average largest entanglement cluster size $\mathbb{E}[S_n^{\max}(G,n^{\star})]$ and examine it in more detail. We consider the regular square and triangular lattices, and we assume that all elementary links have the same success probability $p$. We also consider $n=10$ trials. Then, we find that as the size of the network increases, a critical value of $p$ emerges, call it $p_{\text{crit}}$, at which the average largest entanglement cluster size undergoes a sharp transition. Below $p_{\text{crit}}$, the average largest entanglement cluster size is effectively zero, while beyond $p_{\text{crit}}$ the average largest entanglement cluster size increases to one; see Fig. \ref{fig-avg_cluster_size_all}. We also observe that as $n^{\star}$ increases $p_{\text{crit}}$ decreases; see Table \ref{table-p_crit}.
	
	\begin{table}
		\centering
		\begin{tabular}{|c||c|c|c|c|c|}
			\hline $n^{\star}$ & 0 & 1 & 2 & 3 & 4 \\ \hline\hline
			Square & 0.500 & 0.336 & 0.250 & 0.203 & 0.166 \\ \hline
			Triangular & 0.347 & 0.213 & 0.151 & 0.117 & 0.098  \\ \hline
		\end{tabular}
		\caption{Estimated values of the critical elementary link success probability $p_{\text{crit}}$, based on the curves in Fig. \ref{fig-avg_cluster_size_all}.}\label{table-p_crit}
	\end{table}
	
	The quantity $p_{\text{crit}}$ can be regarded as the minimum elementary link success probability that must be attained in any practical implementation of a large-scale quantum network. In other words, all of the elements that contribute to the elementary link success probability, such as the source inefficiency, the transmission loss, the quantum memory read/write inefficiencies, and the success probability of entanglement purification, have to combine to be greater than $p_{\text{crit}}$ in order to have a good large-scale quantum network. In this context, the values in Fig. \ref{fig-avg_cluster_size_all} imply that the triangular lattice topology is more suitable for large-scale quantum networks since it has a lower critical elementary link success probability for every cutoff value considered.

\section{Summary \& outlook}\label{sec-summary}

	Generation of elementary links is a crucial first step to obtaining long-distance entanglement in a quantum network. In this work, we considered the limitations imposed on quantum networks due to the inherently probabilistic nature of elementary link generation. We proposed the average connection time and the average largest entanglement cluster size as relevant quantities to consider when evaluating the performance of a quantum network. We provided bounds on these two quantities for a particular class of quantum repeater protocols. These bounds led to requirements on the coherence times of quantum memories (Table \ref{tab-n_star_crit}), requirements on the lengths of repeater chains in order to achieve rates that surpass the repeaterless capacity (Table \ref{tab-threshold_length}), and requirements on overall device efficiency for large-scale networks (Table \ref{table-p_crit}).
	
	One direction for future work is to investigate the trade-off between the two quantities considered here and the fidelity of the shared entangled state at the end of the protocol. By considering more general operations at the intermediate nodes, one could then aim to determine quantum repeater protocols that are optimal for these two quantities, similarly to the investigation in Ref. \cite{RST+18} on the trade-off between fidelity and success probability in entanglement purification protocols.
	
	Another direction for future work is to explore how the results obtained here can be generalized to ``one way'' quantum repeater protocols, in which the entanglement to be shared and/or the quantum information to be transmitted is generated entirely locally at a particular node and sent through elementary links to the desired end nodes \cite{GKL+03,RHG05,FWH+10,MSD+12,MKL+14,NJKL16,MLK+16,MEL17}. Such protocols do not require the two-way classical communication that is required in the protocols that we consider; however, these protocols require the use of quantum error-correction codes, which typically results in a significant resource overhead in terms of the number of required physical qubits.
	

\begin{acknowledgments}
S.K. acknowledges support from the National Science Foundation and the National Science and Engineering Research Council of Canada Postgraduate Scholarship. J.P.D., C.T.M., and A.U.S. would like to acknowledge support from the Army Research Office, Air Force Office of Scientific Research, Defense Advanced Research Projects Agency, National Science Foundation and Northrop Grumman Aerospace Systems. We acknowledge valuable discussions with Anthony Brady and Siddhartha Das.
\end{acknowledgments}

\bibliography{kmsd}{}

\appendix

\widetext

\section{Proof of Theorem \texorpdfstring{\ref{thm-dist_time_bounds}}{[]}}\label{app-pf_Nfull_M_nInfty}

	We start by determining the probability distribution of $N(M,\infty)$. Since $N(M,\infty)=\max\{N_1,\dotsc,N_M\}$, we first characterize the set $\mathcal{S}_n\coloneqq \{(n_1,n_2,\dotsc,n_M):\max\{n_1,n_2,\dotsc,n_M\}=n\}$. Observe that $\mathcal{S}_{M,n}$ can be written as the disjoint union $\mathcal{S}_{M,n}=\sqcup_{j=1}^M \mathcal{S}_{M,n}^j$ of the sets $\mathcal{S}_{M,n}^j$, which are given by
	\begin{equation}
		\mathcal{S}_{M,n}^j\coloneqq\bigsqcup_{1\leq k_1<\dotsb <k_j\leq M}\left\{(i_1,i_2,\dotsc,i_M):i_{k_1}=\dotsb=i_{k_j}=n,1\leq i_{\ell}\leq n-1,\ell\notin\{k_1,\dotsc,k_j\}\right\}.
	\end{equation}
	In other words, the set $\mathcal{S}_{M,n}^j$ consists of all $M$-tuples in which $j$ elements of the tuple are equal to $n$ and the rest are between 1 and $n-1$. The largest element of each $M$-tuple is thus equal to $n$, as required. For example, in the case $M=3$, we have
	\begin{align}
		\mathcal{S}_{3,n}^1&=\{(i,j,n):1\leq i,j\leq n-1\}\cup\{(i,n,j):1\leq i,j\leq n-1\}\cup\{(n,i,j):1\leq i,j\leq n-1\},\\
		\mathcal{S}_{3,n}^2&=\{(i,n,n):1\leq i\leq n-1\}\cup\{(n,i,n):1\leq i\leq n-1\}\cup\{(n,n,i):1\leq i\leq n-1\},\\
		\mathcal{S}_{3,n}^3&=\{(n,n,n)\}.
	\end{align}
	
	Since all of the sets $\mathcal{S}_{M,n}^j$ are disjoint, and $\mathcal{S}_{M,n}^j$ is itself a disjoint union of sets, we obtain
	\begin{align}
		\Pr\!\left[N(M,\infty)=n\right]&=\sum_{j=1}^M\Pr\!\left[\mathcal{S}_{M,n}^j\right]\\
		&=\sum_{j=1}^M\sum_{1\leq k_1<\dotsb< k_j\leq M}\sum_{\substack{1\leq i_{\ell}\leq n-1,\\\ell\notin\{k_1,\dotsc,k_j\}}}\Pr\!\left[N_1=i_1,N_2=i_2,\dotsc,N_M=i_M:i_{k_1}=\dotsb=i_{k_j}=n\right]
	\end{align}
	Since all of the random variables $N_i$ are independent, we obtain
	\begin{align}
		\Pr\!\left[N(M,\infty)=n\right]&=\sum_{j=1}^M\sum_{1\leq k_1<\dotsb<k_j\leq M}\prod_{i=1}^j\Pr[N_{k_i}=n]\prod_{\ell\notin\{k_1,\dotsc,k_j\}}\sum_{i_{\ell}=1}^{n-1}\Pr[N_{\ell}=i_{\ell}]
	\end{align}
	By definition, $\Pr[N_{k_i}=n]=p_{k_i}(1-p_{k_i})^{n-1}$, and it is straightforward to show that, if $p_{k_i}=p$ for all $k_i$, then
	\begin{equation}\label{eq-Nfull_M_nInfty_pf}
		\sum_{i=1}^{n-1}(1-p)^{i-1}=\frac{1-(1-p)^{n-1}}{p}.
	\end{equation}
	Therefore, since there are $\binom{M}{j}$ elements in the set $\{(k_1,\dotsc,k_j):1\leq k_1<\dotsb <k_j\leq M\}$, we obtain
	\begin{equation}
		\Pr\!\left[N(M,\infty)=n\right]=\sum_{j=1}^M\binom{M}{j}\left(\frac{1-(1-p)^{n-1}}{p}\right)^{M-j}p^M\left((1-p)^{n-1}\right)^j,
	\end{equation}
	which can be simplified to
	\begin{equation}\label{eq-Pr_Nfull_M_nInfty}
		\Pr\!\left[N(M,\infty)=n\right]=\left(1-(1-p)^{n}\right)^M-\left(1-(1-p)^{n-1}\right)^M.
	\end{equation}
	
	Next, to find $\mathbb{E}\left[N(M,\infty)\right]$, we use the fact that
	\begin{equation}
		\mathbb{E}\left[N(M,\infty)\right]=\sum_{n=1}^{\infty}\Pr\left[N(M,\infty)\geq n\right]=\sum_{n=1}^{\infty}\left(1-\Pr\left[N(M,\infty)<n\right]\right).
	\end{equation}
	Then, since $N(M,\infty)=\max\{N_1,\dotsc,N_M\}$, and since $\max\{N_1,\dotsc,N_M\}<n$ if and only if $N_i<n$ for all $1\leq i\leq M$, we obtain
	\begin{equation}
		\mathbb{E}\left[N(M,\infty)\right]=\sum_{n=1}^{\infty}\left(1-\Pr\left[N_1<n\right]\dotsb\Pr\left[N_M<n\right]\right)=\sum_{n=1}^{\infty}\left(1-\left(1-(1-p)^{n-1}\right)^M\right),
	\end{equation}
	as required, where to obtain the last equality we used Eq. \eqref{eq-Nfull_M_nInfty_pf}, which implies that
	\begin{equation}
		\Pr\left[N_{\ell}<n\right]=\sum_{i=1}^{n-1}p(1-p)^{i-1}=1-(1-p)^{n-1}
	\end{equation}
	for all $1\leq\ell\leq M$. Now,
	\begin{equation}
		\sum_{n=1}^{\infty}\left(1-\left(1-(1-p)^{n-1}\right)^M\right)=\lim_{s\to\infty}\sum_{n=1}^s\left(1-\left(1-(1-p)^{n-1}\right)^M\right).
	\end{equation}
	Letting $q=1-p$, we have
	\begin{equation}
		\left(1-(1-p)^{n-1}\right)^M=\left(1-q^{n-1}\right)^M=\sum_{k=0}^M\binom{M}{k}(-1)^k q^{k(n-1)}=1+\sum_{k=1}^M\binom{M}{k}(-1)^kq^{k(n-1)}.
	\end{equation}
	Therefore,
	\begin{equation}
		\mathbb{E}\left[N(M,\infty)\right]=\lim_{s\to\infty}\sum_{n=1}^s\sum_{k=1}^M\binom{M}{k}(-1)^{k+1}q^{k(n-1)}=\lim_{s\to\infty}\sum_{k=1}^M\binom{M}{k}(-1)^{k+1}\left(\frac{1-q^{ks}}{1-q^k}\right),
	\end{equation}
	where to obtain the last equality we used the fact that
	\begin{equation}
		\sum_{n=1}^s q^{k(n-1)}=\frac{1-q^{ks}}{1-q^k}
	\end{equation}
	for all $k\geq 1$. Finally, for $0\leq q< 1$, it holds that $\lim_{s\to\infty} q^{ks}=0$ for all $k\geq 1$, which means that
	\begin{equation}
		\mathbb{E}\left[N(M,\infty)\right]=\sum_{k=1}^M\binom{M}{k}(-1)^{k+1}\frac{1}{1-q^k}=\sum_{k=1}^M\binom{M}{k}\frac{(-1)^{k+1}}{1-(1-p)^k},
	\end{equation}
	as required.

\section{Proof of Equation \texorpdfstring{\eqref{eq-Pr_NAB_infty}}{} and Equation \texorpdfstring{\eqref{eq-N_AB_nInfty}}{}}\label{app-parallel_link}
	
	We now prove that
	\begin{equation}
		\Pr\left[N(M,\infty;n_P)=n\right]=\left(1-\left(1-(1-p)^{n-1}\right)^M\right)^{n_P}-\left(1-\left(1-(1-p)^n\right)^M\right)^{n_P},
	\end{equation}
	and that the average number of trials needed to connect the $M$ elementary links is
	\begin{equation}
		\mathbb{E}\left[N(M,\infty;n_P)\right]=\sum_{n=1}^{\infty}\left(1-\left(1-(1-p)^{n-1}\right)^M\right)^{n_P}.
	\end{equation}

	In order to prove Eq. \eqref{eq-Pr_NAB_infty} and Eq. \eqref{eq-N_AB_nInfty}, the main task is to characterize the set $\widetilde{\mathcal{S}}_{n_P,n}\coloneqq\{(n_1,n_2,\dotsc,n_{n_P}):\min\{n_1,n_2,\dotsc,n_{n_P}\}=n\}$. It holds that $\widetilde{\mathcal{S}}_{n_P,n}=\sqcup_{j=1}^{n_P}\widetilde{\mathcal{S}}_{n_P,n}^j$, where
	\begin{equation}
		\widetilde{\mathcal{S}}_{n_P,n}^j\coloneqq \bigsqcup_{1\leq k_1<\dotsb<k_j\leq n_P}\left\{(i_1,i_2,\dotsc,i_{n_P}):i_{k_1}=\dotsb=i_{k_j}=n,~i_{\ell}>n,\ell\notin\{k_1,\dotsc,k_j\}\right\}.
	\end{equation}
	Since all of the sets $\widetilde{\mathcal{S}}_{n_P,n}^j$ are disjoint, and $\widetilde{\mathcal{S}}_{n_P,n}^j$ is itself a disjoint union, we obtain
	\begin{align}
		&\Pr\left[N(M,\infty;n_P)=n\right]\nonumber\\
		&\quad=\sum_{j=1}^{n_P}\Pr\left[\widetilde{\mathcal{S}}_{n_P,n}\right]\\
		&\quad=\sum_{j=1}^{n_P}\sum_{1\leq k_1<\dotsb<k_j\leq n_P}\sum_{\substack{i_{\ell}=n+1,\\\ell\notin\{k_1,\dotsc,k_j\}}}^{\infty}\Pr\left[N^1(M,\infty)=i_1,N^2(M,\infty)=i_2,\dotsc,N^{n_P}(M_{n_P},\infty)=i_{n_P}:i_{k_1}=\dotsb=i_{k_j}=n\right]\\
		&\quad=\sum_{j=1}^{n_P}\sum_{1\leq k_1<\dotsb<k_j\leq n_P}\prod_{i=1}^j\Pr\left[N^{k_i}(M,\infty)=n\right]\prod_{\ell\notin\{k_1,\dotsc,k_j\}}\sum_{i_{\ell}=n+1}^{\infty}\Pr\left[N^{\ell}(M,\infty)=i_{\ell}\right].
	\end{align}
	
	Now, let us recall from Eq. \eqref{eq-Pr_Nfull_M_nInfty} that for $n^{\star}=\infty$ we have that
	\begin{equation}
		\Pr\left[N^{\ell}(M,\infty)=i\right]=\left(1-(1-p)^i\right)^{M}-\left(1-(1-p)^{i-1}\right)^{M}
	\end{equation}
	for all $1\leq \ell\leq n_P$. Since
	\begin{equation}\label{eq-N_AB_infty_pf_1}
		\sum_{i_{\ell}=n+1}^{\infty}\Pr\left[N^{\ell}(M,\infty)=i_{\ell}\right]=1-\sum_{i_{\ell}=1}^n\Pr\left[N^{\ell}(M,\infty)=i_{\ell}\right]=1-\left(1-(1-p)^n\right)^{M},
	\end{equation}
	we find that
	\begin{align}
		\Pr\left[N(M,\infty;n_P)=n\right]&=\sum_{j=1}^{n_P}\binom{n_P}{j}\left(\left(1-(1-p)^n\right)^M-\left(1-(1-p)^{n-1}\right)^M\right)^j\left(1-\left(1-(1-p)^n\right)^M\right)^{n_P-j}\\
		&=\left(1-\left(1-(1-p)^{n-1}\right)^M\right)^{n_P}-\left(1-\left(1-(1-p)^n\right)^M\right)^{n_P},
	\end{align}
	as required.
	
	Next, to find $\mathbb{E}\left[N(M,\infty;n_P)\right]$, we use the fact that
	\begin{equation}
		\mathbb{E}\left[N(M,\infty;n_P)\right]=\sum_{n=1}^{\infty}\Pr\left[N(M,\infty;n_P)\geq n\right].
	\end{equation}
	Since $N(M,\infty;n_P)=\min\left\{N^1(M,\infty),\dotsc,N^{n_P}(M,\infty)\right\}$, and since $\min\left\{N^1(M,\infty),\dotsc,N^{n_P}(M,\infty)\right\}\geq n$ if and only if $N^i(M,\infty)\geq n$ for all $1\leq i\leq n_P$, we obtain
	\begin{equation}
		\mathbb{E}\left[N(M,\infty;n_P)\right]=\sum_{n=1}^{\infty}\Pr\left[N^1(M,\infty)\geq n\right]\dotsb\Pr\left[N^{n_P}(M,\infty)\geq n\right]=\sum_{n=1}^{\infty}\left(1-\left(1-(1-p)^{n-1}\right)^M\right)^{n_P},
	\end{equation}
	as required, where to obtain the last equality we made use of the fact that
	\begin{equation}
		\Pr\left[N^i(M,\infty)\geq n\right]=1-\Pr\left[N^i(M,\infty)<n\right]=1-\sum_{j=1}^{n-1}\Pr\left[N^i(M,\infty)=j\right]=1-\left(1-(1-p)^{n-1}\right)^{M}
	\end{equation}
	for all $1\leq i\leq n_P$, which follows from Eq. \eqref{eq-N_AB_infty_pf_1}. This completes the proof.

\section{The average number of established elementary links}\label{app-L_n}

	Given a network with a total of $M$ elementary links and a cutoff of $n^{\star}\geq 0$, the quantity $L_n(M,n^{\star})$ is defined as the number of established elementary links after $n\geq 1$ trials when initially there are no established elementary links in the network. In this section, we provide a general expression for $\mathbb{E}\left[L_n(M,n^{\star})\right]$ and prove Theorem \ref{thm-L_n_bounds}.
	
	Let us start by defining $L^{(j)}(M,n^{\star})$ to be the number of elementary links established in the $j^{\text{th}}$ trial, where $j\geq 1$. In the case $j=1$, we have $0\leq L^{(1)}(M,n^{\star})\leq M$. For $1<j\leq n^{\star}+1$, none of the established elementary links are reset between trials. This means that $L^{(j)}(M,n^{\star})$ depends on $L^{(1)}(M,n^{\star}), L^{(2)}(M,n^{\star}),\dotsc,L^{(j-1)}(M,n^{\star})$. If $x_i$ represents the number established elementary links in the $i^{\text{th}}$ trial, then
	\begin{equation}\label{eq-avg_num_links_j_range}
		0\leq L^{(j)}(M,n^{\star})\leq M-x_1-x_2-\dotsb-x_{j-1}, \quad 1<j\leq n^{\star}+1.
	\end{equation}
	For $j>n^{\star}+1$, elementary links start being reset: before the start of trial $j=n^{\star}+2$, all links established in the first trial are reset, which means that $0\leq L^{(n^{\star}+2)}(M,n^{\star})\leq M-x_2-\dotsb -x_{n^{\star}+1}$. Then, before the start of trial $j=n^{\star}+3$, all links established in the second trial are reset, which means that $0\leq L^{(n^{\star}+3)}(M,n^{\star})\leq M-x_3-\dotsb -x_{n^{\star}+2}$. In general, then, $L^{(j)}(M,n^{\star})$ depends on $L^{(j-n^{\star})}(M,n^{\star}),\dotsc,L^{(j-2)}(M,n^{\star}),L^{(j-1)}(M,n^{\star})$, which means that
	\begin{equation}
		0\leq L^{(j)}(M,n^{\star})\leq M-x_{j-n^{\star}}-\dotsb-x_{j-2}-x_{j-1},\quad j>n^{\star}+1.
	\end{equation}
	
	Let us now consider the probability distribution of the random variables $L^{(j)}(M,n^{\star})$. First, for $j=1$, we have
	\begin{equation}\label{eq-avg_num_links_j1_pr}
		\Pr\left[L^{(1)}(M,n^{\star})=x\right]=\binom{M}{x}p^{x}(1-p)^{M-x},\quad 0\leq x\leq M.
	\end{equation}
	
	For all $j\leq n^{\star}+1$, because none of the links are reset between trials, we have that $L^{(j)}(M,n^{\star})$ depends on all trials before the $j^{\text{th}}$ one, so that
	\begin{multline}
		\Pr\left[L^{(j)}(M,n^{\star})=x~\big\vert~L^{(1)}(M,n^{\star})=x_1,L^{(2)}(M,n^{\star})=x_2,\dotsc,L^{(j-1)}(M,n^{\star})=x_{j-1}\right]=\\ \binom{M-x_1-x_2-\dotsb-x_{j-1}}{x}p^x(1-p)^{M-x_1-x_2-\dotsb-x_{j-1}}
	\end{multline}
	for all $2\leq j\leq n^{\star}+1$ and all $0\leq x\leq M-x_1-x_2-\dotsb-x_{j-1}$. For trials $j>n^{\star}+1$, elementary links start being reset as described above. This means that $L^{(j)}(M,n^{\star})$ depends only on the $n^{\star}$ trials prior to the $j^{\text{th}}$ trial, i.e.,
	\begin{align*}
		&\Pr\left[L^{(j)}(M,n^{\star})=x~\big\vert~L^{(1)}(M,n^{\star})=x_1,L^{(2)}(M,n^{\star})=x_2,\dotsc,L^{(j-1)}(M,n^{\star})=x_{j-1}\right] \\
		&\qquad=\Pr\left[L^{(j)}(M,n^{\star})=x~\big\vert~L^{(j-n^{\star})}(M,n^{\star})=x_{j-n^{\star}},\dotsc,L^{(j-2)}(M,n^{\star})=x_{j-2},L^{(j-1)}(M,n^{\star})=x_{j-1}\right]\\
		&\qquad=\binom{M-x_{j-n^{\star}}-\dotsb-x_{j-2}-x_{j-1}}{x}p^x(1-p)^{M-x_{j-n^{\star}}-\dotsb-x_{j-2}-x_{j-1}-x}
	\end{align*}
	for all $j>n^{\star}+1$ and all $0\leq x\leq M-x_{j-n^{\star}}-\dotsb-x_{j-2}-x_{j-1}$.
	
	
	Let us now consider the quantity $L_n(M,n^{\star})$. In the case $n^{\star}=\infty$, once an elementary link has been established, it never has to be reset. This implies that $L_n(M,\infty)=\sum_{j=1}^n L^{(j)}(M,\infty)$. This equality holds even for finite $n^{\star}$, provided that $n\leq n^{\star}+1$, because in this case none of the established elementary links have to be reset (because the cutoff $n^{\star}$ is never reached). This means that $L_n(M,n^{\star})=\sum_{j=1}^n L^{(j)}(M,n^{\star})$ for all $0\leq n^{\star}\leq \infty$ and all $n\leq n^{\star}+1$. If $n>n^{\star}+1$, then every established elementary link is reset $n^{\star}$ trials after it was established. This means that the number of elementary links established in the $j^{\text{th}}$ trial must be removed from the total number of established elementary links $n^{\star}$ trials after the $j^{\text{th}}$ trial. Therefore,
	\begin{equation}
		L_n(M,n^{\star})=\sum_{j=1}^n L^{(j)}(M,n^{\star})-\sum_{j=n^{\star}+1}^{n-1} L^{(j-n^{\star})}(M,n^{\star})=\sum_{j=n-n^{\star}}^n L^{(j)}(M,n^{\star}),\quad n>n^{\star}+1.
	\end{equation}
	In other words, for $n>n^{\star}+1$, only the last $n^{\star}+1$ trials matter for determining the total number of established elementary links after $n$ trials. In summary, for all $0\leq n^{\star}\leq \infty$ and all $n\geq 1$,
	\begin{equation}
		L_n(M,n^{\star})=\left\{\begin{array}{l l} \displaystyle \sum_{j=1}^n L^{(j)}(M,n^{\star}), & n\leq n^{\star}+1,\\[0.1cm]
		\displaystyle \sum_{j=n-n^{\star}}^n L^{(j)}(M,n^{\star}), & n>n^{\star}+1. \end{array}\right.
	\end{equation}
	Note that
	\begin{equation}\label{eq-avg_num_links_0}
		L_n(M,0)=L^{(1)}(M,0),
	\end{equation}
	which means that the number of established elementary links in $n$ trials with $n^{\star}=0$ is the same as the number of established elementary links after one trial. This makes sense, since for $n^{\star}=0$ all established elementary links are reset at the end of each trial.

\subsection{Proof of Theorem \texorpdfstring{\ref{thm-L_n_bounds}}{[]}}\label{app-L_n_infty}

	We now prove that
	\begin{equation}\label{eq-avg_num_links_0_2}
		\frac{1}{M}\mathbb{E}\left[L_n(M,0)\right]=p,\quad n\geq 1,
	\end{equation}
	and
	\begin{equation}\label{eq-avg_num_links_infty_2}
		\frac{1}{M}\mathbb{E}\left[L_n(M,\infty)\right]=1-(1-p)^n,\quad n\geq 1.
	\end{equation}
	The latter can be stated more generally as
	\begin{equation}\label{eq-avg_num_links_infty_3}
		\frac{1}{M}\mathbb{E}\left[L_n(M,n^{\star})\right]=1-(1-p)^n,\quad n\leq n^{\star}+1.
	\end{equation}
	
	Since by Eq. \eqref{eq-avg_num_links_0} we have that $L_n(M,0)=L^{(1)}(M,0)$, and we have from Eq. \eqref{eq-avg_num_links_j1_pr} that $L^{(1)}(M,0)$ is simply a binomial random variable, we immediately obtain $\mathbb{E}\left[L_n(M,0)\right]=\mathbb{E}\left[L^{(1)}(M,0)\right]=Mp$, so that Eq. \eqref{eq-avg_num_links_0_2} holds.
	
	To prove Eq. \eqref{eq-avg_num_links_infty_2}, and more generally Eq. \eqref{eq-avg_num_links_infty_3}, we first calculate $\mathbb{E}\left[L^{(j)}(M,n^{\star})\right]$ with $1\leq j\leq n^{\star}+1$. Using Eq. \eqref{eq-avg_num_links_j_range}, we have
	\begin{align}
		&\Pr\left[L^{(j)}(M,n^{\star})=x\right]\nonumber\\
		&\quad=\sum_{x_1=0}^M\sum_{x_2=0}^{M-x_1}\dotsb\sum_{x_{j-1}=0}^{M-x_1-x_2-\dotsb-x_{j-2}}\Pr\left[L^{(1)}(M,n^{\star})=x_1,L^{(2)}(M,n^{\star})=x_2,\dotsc,L^{(j-1)}(M,n^{\star})=x_{j-1},L^{(j)}(M,n^{\star})=x\right]\\
		&\quad=\sum_{x_1=0}^{M}\sum_{x_2=0}^{M-x_1}\dotsb\sum_{x_{j-1}=0}^{M-x_1-x_2-\dotsb-x_{j-2}}\Pr\left[L^{(1)}(M,n^{\star})=x_1\right]\Pr\left[L^{(2)}(M,n^{\star})=x_2~\big\vert~L^{(1)}(M,n^{\star})=x_1\right]\times\nonumber\\
		&\quad\qquad\dotsb\times \Pr\left[L^{(j)}(M,n^{\star})=x~\big\vert~L^{(1)}(M,n^{\star})=x_1,L^{(2)}(M,n^{\star})=x_2,\dotsc,L^{(j-1)}(M,n^{\star})=x_{j-1}\right].
	\end{align}
	Therefore,
	\begin{align}
		\mathbb{E}\left[L^{(j)}(M,n^{\star})\right]&=\sum_{x=0}^{M-x_1-x_2-\dotsb-x_{j-1}}x\Pr\left[L^{(j)}(M,n^{\star})=x\right]\\
		&=\sum_{x_1=0}^{M}\sum_{x_2=0}^{M-x_1}\dotsb\sum_{x_{j-1}=0}^{M-x_1-x_2-\dotsb-x_{j-2}}x\Pr\left[L^{(1)}(M,n^{\star})=x_1\right]\Pr\left[L^{(2)}(M,n^{\star})=x_2~\big\vert~L^{(1)}(M,n^{\star})=x_1\right]\times\nonumber\\
		&\qquad\dotsb\times \Pr\left[L^{(j)}(M,n^{\star})=x~\big\vert~L^{(1)}(M,n^{\star})=x_1,L^{(2)}(M,n^{\star})=x_2,\dotsc,L^{(j-1)}(M,n^{\star})=x_{j-1}\right].
	\end{align}
	
	Now,
	\begin{equation}
		\sum_{x=0}^{M-x_1-x_2-\dotsb-x_{j-1}}x\binom{M-x_1-x_2-\dotsb-x_{j-1}}{x}p^x(1-p)^{M-x_1-x_2-\dotsb-x_{j-1}-x}=(M-x_1-x_2-\dotsb-x_{j-1})p.
	\end{equation}
	Then,
	\begin{align}
		&\sum_{x_{j-1}=0}^{M-x_1-x_2-\dotsb-x_{j-2}}\binom{M-x_1-x_2-\dotsb-x_{j-2}}{x_{j-1}}p^{x_{j-1}}(1-p)^{M-x_1-x_2-\dotsb-x_{j-2}-x_{j-1}}(M-x_1-x_2-\dotsb-x_{j-2}-x_{j-1})\nonumber\\
		&\qquad\qquad = M-x_1-x_2-\dotsb-x_{j-2}-(M-x_1-x_2-\dotsb-x_{j-2})p\\
		&\qquad\qquad = (M-x_1-x_2-\dotsb-x_{j-2})(1-p).
	\end{align}
	Similarly, summing over $x_{j-2}$ gives the result $(M-x_1-x_2-\dotsb-x_{j-3})(1-p)$. Continuing this for all the summation variables, we ultimately obtain
	\begin{equation}
		\mathbb{E}\left[L^{(j)}(M,n^{\star})\right]=Mp(1-p)^{j-1},\quad n^{\star}\geq 1,\quad 1\leq j\leq n^{\star}+1.
	\end{equation}
	Using this, we find that for all $n\leq n^{\star}+1$,
	\begin{equation}
		\mathbb{E}\left[L_n(M,n^{\star})\right]=\sum_{j=1}^n\mathbb{E}\left[L^{(j)}(M,n^{\star})\right]=\sum_{j=1}^n Mp(1-p)^{j-1}=M(1-(1-p)^n),
	\end{equation}
	as required.

\end{document}